\documentclass[pra,twocolumn,superscriptaddress,showpacs,preprintnumbers,amsmath,amssymb]{revtex4}
\usepackage{mathrsfs}
\usepackage{bbm}
\usepackage{amsfonts}
\usepackage{tipa}

\usepackage{epsfig,graphicx}
\usepackage{amstext}
\usepackage{amsmath}
\usepackage{graphicx}
\usepackage{times}
\usepackage{graphicx}
\usepackage{times}

\begin{document}


\title{Enhancing the geometric quantum discord in the Heisenberg {\it XX} chain
       by Dzyaloshinsky-Moriya interaction}
\author{Jia-Min Gong}
\email{jmgong@yeah.net}
\address{School of Electronic Engineering, Xi'an University of Posts and
         Telecommunications, Xi'an 710121, China}
\author{Qi Tang}
\address{School of Electronic Engineering, Xi'an University of Posts and
         Telecommunications, Xi'an 710121, China}
\author{Yu-Hang Sun}
\address{School of Science, Xi'an University of Posts and
         Telecommunications, Xi'an 710121, China}
\author{Lin Qiao}
\address{School of Electronic Engineering, Xi'an University of Posts and
         Telecommunications, Xi'an 710121, China}

\begin{abstract}
We studied the trance distance, the Hellinger distance, and the Bures
distance geometric quantum discords (GQDs) for a two-spin Heisenberg
{\it XX} chain with the Dzyaloshinsky-Moriya (DM) interaction and the
external magnetic fields. We found that considerable enhancement of the
GQDs can be achieved by introducing the DM interaction, and their
maxima were obtained in the limiting case $D\rightarrow \infty$. The
external magnetic fields and the increase of the temperature can also
enhance the GQDs to some extent for certain special cases.

\end{abstract}

\maketitle

\section{Introduction}\label{sec:1}
For a long time, entanglement was considered to be the only resource
responsible for the advantage of many quantum information processing
(QIP) tasks \cite{rmp-e,qt1,qt2,disen}. As entanglement exists only
in the non-separable states, separable states were also considered to
be classically correlated and useless for QIP. But recent studies
revealed that the separable states may also possess certain kinds of
quantum correlations. For example, the quantum discord (QD)
\cite{qd-en}, which is a more general quantum correlation measure
than that of entanglement, can be nonzero for some separable states
\cite{rmp-qd}. From a practical point of view, it is proposed that
the QD is responsible for the power of the QIP tasks such as the
deterministic quantum computation with one qubit \cite{dqc1},
remote state preparation \cite{rsp}, and quantum locking
\cite{qlock1,qlock2}. The QD is also intimately related to many
fundamental problems of quantum mechanics \cite{Gumile,cpm,eur1,eur2}.

Besides the entropic measure of QD \cite{qd-en}, the quantumness of a
state can also be characterized from many other perspectives. These
measures include the measurement-induced disturbance \cite{mid}, and
the measurement-induced nonlocality \cite{min1,min2,min3}. Also there
are distance-based quantum correlation measures, such as the first
proposed geometric QD (GQD) defined via the Hilbert-Schmidt distance
\cite{gqd-sch} (which may be changed by trivial local actions on the
unmeasured party \cite{Problem}), and its modified version via the
trace distance \cite{trace} or the Hellinger distance \cite{square,lqu}.
Moreover, the GQD in a state can also be defined via the Bures distance
\cite{bures}.

The above progress prompted a huge surge of people's interest in this
new field. Particularly, as a potential and costly resource for QIP,
the long-time preservation of QDs remains a main pursuit of people
\cite{pres1,pres2,pres3}, and their decay dynamics for various open
quantum systems have been studied, with many novel phenomena being
observed \cite{dyn1,dyn2,dyn3,dyn4,dyn5,dyn6,dyn7,dyn-trace}. The QDs
in various spin systems \cite{sp1,sp2,sp3,sp4,sp5,sp6,sp7,sp8}, and
its role in detecting quantum phase transition points at finite
temperatures have also been revealed \cite{qpt,qpt1,qpt2,qpt3}.

We investigate in this work the properties of a two-spin system
described by the Heisenberg {\it XX} model. Different from those
previous studies, we introduced here the Dzyaloshinsky-Moriya (DM)
interaction induced by the spin-orbit coupling, whose effects on the
properties of entanglement \cite{DM1,DM2} and the entropic QD
\cite{sp8} have already been studied. Here, we concentrate on its
effects on the trace distance, the Hellinger distance, and the
Bures distance GQDs, which have been demonstrated to be well-defined
measures of quantum correlations. We will show that for the
considered physical model, the three GQDs can be enhanced obviously by
increasing the strength of the DM interaction.

This paper is arranged as follows. In Section \ref{sec:2} we recall
the definitions for three different distanced-based quantum
correlation measures we adopted. In Section \ref{sec:3}, we
give the physical model we considered, and some analytical results
obtained for the three GQDs. Section \ref{sec:4} is devoted to a
discussion of the dependence of the GQDs on the system parameters,
through which we show that the GQDs can be enhanced obviously by
increasing the DM interaction. We conclude this paper in Section
\ref{sec:5}.

\section{Basic formalism for the GQD{\sc s}}\label{sec:2}
We recall in this section the definitions and the related
formula for three types of the GQDs we adopted in this paper, namely,
the trace distance, the Hellinger distance, and the Bures distance
GQDs \cite{trace,square,lqu,bures}. They characterize the quantum
correlations of a bipartite state $AB$ with the density operator
$\rho$ from different perspectives, and can be classified as the
distance-based measures of quantum correlations as they are all
related to certain forms of distance.

First, we recall the trace distance GQD for a bipartite state
$\rho$, which is defined as \cite{trace}
\begin{eqnarray}\label{eq1}
 Q_T(\rho)=\min_{\chi\in\rho_{CQ}}||\rho-\chi||_1,
\end{eqnarray}
where $||\rho-\chi||_1={\rm Tr}\sqrt{(\rho-\chi)^\dag (\rho-\chi)}$
denotes the trace distance between $\rho$ and $\chi\in\rho_{CQ}$,
and
\begin{eqnarray}\label{eq2}
 \rho_{CQ}=\sum_k p_k \Pi_k^A\otimes \rho_k^B,
\end{eqnarray}
is the classical-quantum state \cite{gqd-sch}, where $\{p_k\}$ is a probability
distribution, and $\Pi_k^A$ and $\rho_k^B$ are the orthogonal
projector for $A$ and the density operator for $B$, respectively.

For the two-qubit {\it X} state $\rho^X$ which only contains nonzero
elements along the main diagonal and anti-diagonal, the calculation
of the trace distance GQD can be simplified,
with the analytical expression being given by \cite{ana-trace}
\begin{small}
\begin{eqnarray}\label{eq3}
 Q_T(\rho^X)=\sqrt{\frac{\gamma_1^2 \max\{\gamma_3^2,\gamma_2^2+x_3^2\}
              -\gamma_2^2\min\{\gamma_1^2,\gamma_3^2\}}
              {\max\{\gamma_3^2,\gamma_2^2+x_3^2\}-\min\{\gamma_1^2,\gamma_3^2\}
              +\gamma_1^2-\gamma_2^2}},
\end{eqnarray}
\end{small}
where $\gamma_{1,2}=2(|\rho^X_{23}|\pm |\rho^X_{14}|)$,
$\gamma_3=1-2(\rho^X_{22}+\rho^X_{33})$, and
$x_3=2(\rho^X_{11}+\rho^X_{22})-1$.

Second, we recall the Hellinger distance GQD. It is defined based
on the square root of the density operator $\rho$, and can be written as \cite{square}
\begin{eqnarray}\label{eq4}
 Q_H(\rho)=2\inf_{\Pi^A}||\sqrt{\rho}-\Pi^A(\sqrt{\rho})||_2^2,
\end{eqnarray}
where the infimum is taken over all the projection-valued
measurements $\Pi^A=\{\Pi_k^A\}$, with $\Pi^A(\sqrt{\rho})=\sum_k
\Pi_k^A\sqrt{\rho}\Pi_k^A$, and $||X||_2=\sqrt{{\rm tr}X^\dagger X}$
is the Hilbert-Schmidt distance.

The calculation of $Q_H(\rho)$ is difficult in general. But for the
special case of $2\times n$ system, $Q_H(\rho)$ can be calculated
via \cite{lqu}
\begin{eqnarray}\label{eq5}
 Q_H(\rho)=1-\lambda_{\max}\{W_{AB}\},
\end{eqnarray}
where $\lambda_{\max}$ denotes the maximum eigenvalue of a $3\times 3$ matrix
$W_{AB}$, whose elements are given by
\begin{eqnarray}\label{eq6}
 (W_{AB})_{ij}={\rm Tr}\{\sqrt{\rho}(\sigma_A^i\otimes I_n)\sqrt{\rho}(\sigma_B^j\otimes I_n)\},
\end{eqnarray}
with $\sigma_S^{x,y,z}$ ($S=A,B$) the three Pauli operators, and $I_n$ the
$n\times n$ identity operator.

Finally, the Bures distance GQD we considered is defined via the Bures
distance between $\rho$ and $\chi\in\rho_{CQ}$ \cite{bures}, which
is of the following form
\begin{eqnarray}\label{eq7}
 Q_B(\rho)=\sqrt{(2+\sqrt{2})\left(1-\sqrt{\max_{\chi\in\rho_{CQ}}F(\rho,\chi)}\right)},
\end{eqnarray}
where $F(\rho,\chi)=[{\rm Tr}(\sqrt{\rho}\chi\sqrt{\rho})^{1/2}]^2$
is the Uhlmann fidelity, and $Q_ B(\rho)$ in Eq. \eqref{eq7} is
normalized, namely, it takes the value 1 for the maximally discordant
states.

For the case of $2\times n$ system, the maximum of $F(\rho,\chi)$
can be calculated via \cite{ana-bures}
\begin{eqnarray}\label{eq8}
 \max_{\chi\in\rho_{CQ}}F(\rho,\chi)=\frac{1}{2}\max_{||\vec{u}=1||}
 \left(1-{\rm tr}\Lambda+2\sum_{k=1}^{n}\lambda_k(\Lambda)\right),
\end{eqnarray}
where $\Lambda=\sqrt{\rho}(\vec{u}\cdot\vec{\sigma}\otimes I_n)
\sqrt{\rho}$, with $\vec{\sigma}=\{\sigma^x,\sigma^y,\sigma^z\}$
being the vector of the Pauli operators, $\vec{u}$ is a unit vector
in $\mathbb{R}^3$, and $\lambda_k(\Lambda)$ denote the eigenvalues
of $\Lambda$ in non-increasing order.

\section{The model}\label{sec:3}
We consider in this work two spins described by the Heisenberg {\it XX}
model, with the addition of the DM interaction which arises from the
spin-orbit coupling being involved. The
corresponding Hamiltonian is given by
\begin{equation}\label{eq9}
 \hat{H}=J (\sigma_1^x\sigma_2^x+\sigma_1^y\sigma_2^y)+B(\sigma_1^z+\sigma_2^z)
 +D(\sigma_1^x\sigma_2^y-\sigma_1^y\sigma_2^x),
\end{equation}
where $\sigma_n^\alpha$ ($\alpha=x,y,z$) are the familiar Pauli
operators acting on the $n$-th spin, $J$ is the coupling constant
between the two spins, while $B$ and $D$ denote respectively the
strengths of the external magnetic field and the DM interaction,
both of which along the $z$ direction. Moreover, $\hbar=1$ is assumed here.

For the physical model described in Eq. \eqref{eq9}, its eigenvalues
and eigenvectors can be derived analytically, which are given
respectively by
\begin{eqnarray}\label{eq10}
\epsilon_{1,2}=\pm 2\delta,~\epsilon_{3,4}=\pm 2B,
\end{eqnarray}
and
\begin{eqnarray}\label{eq11}
 && |\Psi_{1,2}\rangle=\frac{1}{\sqrt{2}}(|10\rangle\pm e^{i\theta}|01\rangle),\nonumber\\
 && |\Psi_3\rangle = |00\rangle,~ |\Psi_4\rangle = |11\rangle.
\end{eqnarray}
where $\delta=\sqrt{J^2+D^2}$, and $\theta=\arctan(D/J)$.

From the above two equations, one can obtain the state of the
system at thermal equilibrium with temperature $T$, which is given by the
Gibb's density operator $\rho=Z^{-1}\exp(-\beta\hat{H})$,
with $\beta=1/k_B T$, and $k_B$ the Boltzman's constant. Moreover,
$Z=\text{Tr}[\exp(-\beta\hat{H})]$ denotes the partition function,
which can be obtained explicitly as
\begin{equation}\label{eq12}
 Z=2(\cosh 2\beta\delta+\cosh 2\beta B),
\end{equation}
while the density operator $\rho$ is given by
\begin{small}
\begin{eqnarray}\label{eq13}
\rho=\frac{1}{Z}
   \left(\begin{array}{cccc}
            e^{-2\beta B}  & 0 & 0 & 0\\
            0  & \cosh 2\beta\delta & -e^{i\theta}\sinh 2\beta\delta & 0 \\
            0  &-e^{-i\theta}\sinh 2\beta\delta & \cosh 2\beta\delta & 0 \\
            0  & 0 & 0 & e^{2\beta B}
  \end{array}\right).
\end{eqnarray}
\end{small}

Clearly, $\rho$ expressed in Eq. \eqref{eq13} is of the X form,
and therefore the trace distance GQD can be derived analytically
as
\begin{equation}\label{eq14}
 Q_T(\rho)=\frac{2\sinh 2\beta\delta}{Z}.
\end{equation}

Moreover, as the square root of the density operator $\rho$ is given
by
\begin{small}
\begin{eqnarray}\label{eq15}
\sqrt{\rho}=\frac{1}{\sqrt{Z}}
   \left(\begin{array}{cccc}
            e^{-\beta B}  & 0 & 0 & 0\\
            0  & 2\cosh \beta\delta & -2e^{i\theta}\sinh \beta\delta & 0 \\
            0  &-2e^{-i\theta}\sinh \beta\delta & 2\cosh \beta\delta & 0 \\
            0  & 0 & 0 & e^{\beta B}
  \end{array}\right),
\end{eqnarray}
\end{small}
the Hellinger distance GQD can be derived analytically as
\begin{equation}\label{eq16}
 Q_H(\rho)=1-{\rm max}\{\lambda_1,\lambda_2\},
\end{equation}
where
\begin{eqnarray}\label{eq17}
 &&\lambda_{1}=\frac{8}{Z}\cosh\beta\delta\cosh\beta B,\nonumber\\
 &&\lambda_2=\frac{1}{Z}(8+2\cosh{2\beta B}).
\end{eqnarray}

Finally, when considering the Bures distance GQD, there
is no analytical expression can be obtained, and thus we calculate
it via numerical methods.

\section{GQDs in the XX chain with DM interaction}\label{sec:4}
We discuss in this section the effects of the DM interaction, the
external magnetic fields, and the reservoir temperature on the
considered GQDs, i.e., the trace distance, the Hellinger distance,
and the Bures distance GQDs introduced in Sec. \ref{sec:2}. As
$Q_T(\rho)$, $Q_B(\rho)$, and $Q_H(\rho)$ are all symmetric
functions with respect to $J=0$, $B=0$, and $D=0$, we consider
in the following only the cases of $J\geq 0$, $B\geq 0$,
and $D\geq 0$.

We begin with a heuristic analysis of the dependence of the
GQDs on $D$ and $B$ for two special cases. First,
for the zero absolute temperature case, the system is in its
ground state, which is given by
$|\Psi_2\rangle$ if $\delta>B$, $|\Psi_4\rangle$ if $\delta<B$, and
the equal mixture of $|\Psi_2\rangle$ and $|\Psi_4\rangle$ if
$\delta=B$. Therefore, we have
\begin{equation}\label{eq18}
  Q_{\alpha}(\rho)=\left\{
    \begin{aligned}
         & 0  &~\text{if}~\delta<B,\\
         & c  &~\text{if}~\delta=B,\\
         & 1  &~\text{if}~\delta>B,
    \end{aligned} \right.
\end{equation}
where $c=0.5$ for $\alpha=\{T, H\}$, and $c\simeq 0.5098$ for
$\alpha=B$.

Second, for the limiting case of $D\rightarrow\infty$, we have
$\rho_{11,44}=0$, $\rho_{22,33}=0.5$,
$\rho_{23}=\rho_{32}^\dag=-0.5i$, and therefore
\begin{equation}\label{eq19}
  Q_{\alpha}(\rho)=1,
\end{equation}
for $\alpha\in\{T,H,B\}$, i.e., they all achieved their maxima in the
limit of $D\rightarrow\infty$. For finite but large enough DM interaction
$D$, it is also natural to hope that one can achieve considerable
enhancement of the three GQDs. We show in the following that
this is indeed the case. Moreover, we remark here that $T$ in all the following figures
is plotted in units of the Boltzman's constant $k_B$.

\subsection{$D$ dependence of the GQDs}

\begin{figure}
\centering
\resizebox{0.36\textwidth}{!}{%
\includegraphics{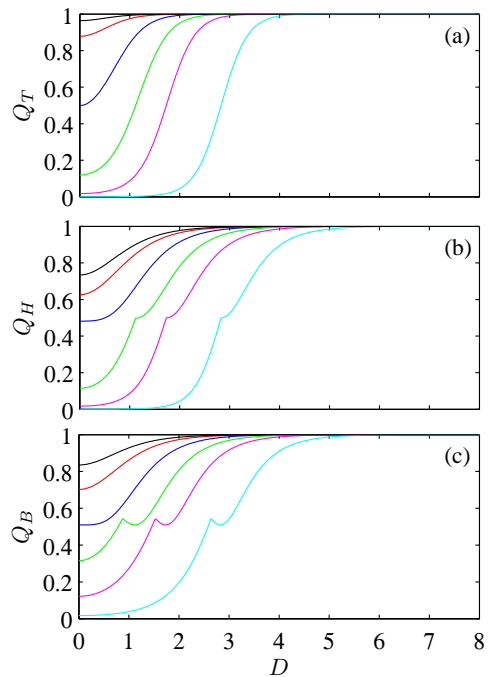}}
\caption{(Color online) $Q_T(\rho)$, $Q_H(\rho)$, and $Q_B(\rho)$
versus $D$ with $J=1$ and $T=0.5$. Here, the black, red,
blue, green, magenta, and cyan curves (from top to bottom)
correspond to the cases of $B=0$, 0.5, 1, 1.5, 2, and 3,
respectively.}\label{fig:1}
\end{figure}

We now discuss the general behaviors of the GQDs at
finite temperature $T$. We first consider their dependence on the DM
interaction. Our numerical simulations show that $Q_T(\rho)$, $Q_H(\rho)$,
and $Q_B(\rho)$ display qualitatively the similar behaviors under different
temperatures. In Fig. \ref{fig:1}, we presented an
exemplified plot of their dependence on $D$ at finite temperature
$T=0.5$ with different strengths of the external magnetic fields. From this plot,
one can note that for any given magnitude of $B$, the trace distance
GQD $Q_T(\rho)$ always increases monotonously with the increasing
value of $D$, and this can be understood from Eq. \eqref{eq14}, which
yields
\begin{equation}\label{eq20}
 \frac{\partial Q_T(\rho)}{\partial\delta}=\frac{8\beta}{Z^2}(1+
                    \cosh 2\beta\delta\cosh 2\beta B)>0.
\end{equation}

Similarly, the Hellinger distance GQD $Q_H(\rho)$ can also be
increased monotonously by increasing $D$. But as displayed in Fig.
\ref{fig:1}(b), it exhibits sudden change behaviors for strong
magnetic fields, and this is caused by the process of maximization
in Eq. \eqref{eq16}, as before the sudden change point denoted by $D_c$, we have
${\rm max}\{\lambda_1,\lambda_2\}=\lambda_2$ and $Q_H(\rho)=1-\lambda_2$, while after $D_c$,
${\rm max}\{\lambda_1,\lambda_2\}=\lambda_{1}$ and $Q_H(\rho)=1-\lambda_1$.

The Bures distance GQD $Q_B(\rho)$ for the relative weak
external magnetic fields case also increases monotonously with the
increase of $D$. But when an strong magnetic field is applied,
$Q_B(\rho)$ exhibits sudden change behavior at a critical point $D_{c1}$.
Particularly, different from that of $Q_H(\rho)$, $Q_B(\rho)$ is
decreased after $D_{c1}$ [see, Fig. \ref{fig:1}(c)], and
this tendency of decrease will continue until another critical point
$D_{c2}$ is reached, after which it turns out to be increased again
and finally approaches the asymptotic value 1.

\subsection{$B$ dependence of the GQDs}

\begin{figure}
\centering
\resizebox{0.36\textwidth}{!}{%
\includegraphics{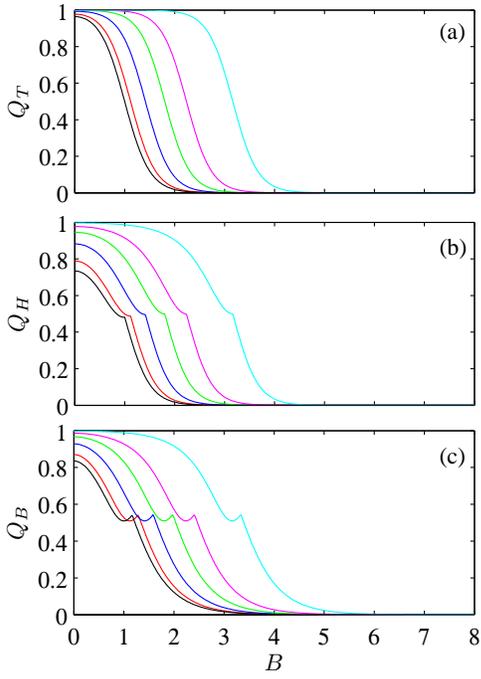}}
\caption{(Color online) $Q_T(\rho)$, $Q_H(\rho)$, and $Q_B(\rho)$
versus $B$ with $J=1$ and $T=0.5$. Here, the black, red, blue,
green, magenta, and cyan curves (from left to right) correspond to
the cases of $D=0$, 0.5, 1, 1.5, 2, and 3,
respectively.}\label{fig:2}
\end{figure}

\begin{figure}
\centering
\resizebox{0.36\textwidth}{!}{%
\includegraphics{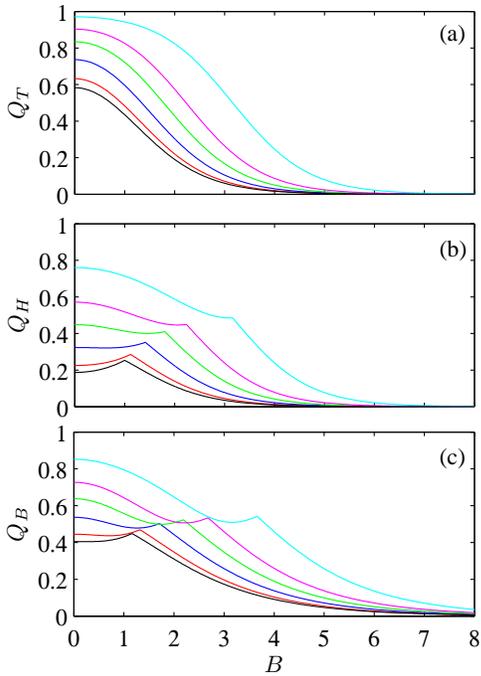}}
\caption{(Color online) $Q_T(\rho)$, $Q_H(\rho)$, and $Q_B(\rho)$
versus $B$ with $J=1$ and $T=1.5$. Here, the black, red, blue,
green, magenta, and cyan curves (from bottom to top) correspond to
the cases of $D=0$, 0.5, 1, 1.5, 2, and 3,
respectively.}\label{fig:3}
\end{figure}

In Figs. \ref{fig:2} and \ref{fig:3}, we displayed the $B$
dependence of the three GQDs under different DM
interactions, with the reservoir temperatures being given by $T=0.5$
and $T=1.5$, respectively. At a first glance, one can note that they
all approach the asymptotic value 0 in the limit of
$B\rightarrow\infty$. This happens because for this special case,
the density operator reduces to $\rho=|11\rangle\langle 11|$, which
is a product state and therefore there is no correlations between
$A$ and $B$.

Another general behavior which can be observed from Figs.
\ref{fig:2} and \ref{fig:3} is that the trace distance GQD
$Q_T(\rho)$ always decreases with the increasing strength of the
external magnetic fields $B$. This phenomenon is obvious because
from Eq. \eqref{eq14} one can see that the partition function $Z$ is
an increasing function of $B$, and therefore $Q_T(\rho)$ always
decreases with the increase of $B$.

The Hellinger distance GQD $Q_H(\rho)$ behaves quite differently
from that of $Q_T(\rho)$. First, there are sudden change behaviors
which are caused by the maximization process appeared in Eq.
\eqref{eq16}, and second, its change with the variation of $B$ is
temperature dependent. As displayed in Figs. \ref{fig:2}(b) and
\ref{fig:3}(b), $Q_H(\rho)$ always decays with the increasing strength
of $B$ at relative low temperature region, while at high temperature region with
weak DM interactions, $Q_H(\rho)$ may be increased by increasing $B$
before the sudden change point $B_c$. But after $B_c$, $Q_H(\rho)$
still decays to zero monotonously.

When considering the Bures distance GQD, as
displayed in Fig. \ref{fig:2}(c) which depicts the relative
low temperature case, $Q_B(\rho)$ is increased with increasing $B$
only in a very narrow region of $B$, while out of this region, it
decays monotonously with $B$. For the case of high temperature with
relative weak DM interaction, as displayed in Fig. \ref{fig:3}(c),
$Q_B(\rho)$ is increased before a sudden change point $B_c$, and
decreased after $B_c$, and this phenomenon is somewhat similar to
that of $Q_H(\rho)$. But when one enlarges the DM interaction, the
$B$ dependence of $Q_B(\rho)$ is changed. It initially decreases
to a minimum value, and then turns out to be increased before a
sudden change point is arrived, after which
it becomes decreasing with $B$ again, and approaches zero in the
infinite $B$ limit.

\subsection{$T$ dependence of the GQDs}

\begin{figure}
\centering
\resizebox{0.36\textwidth}{!}{%
\includegraphics{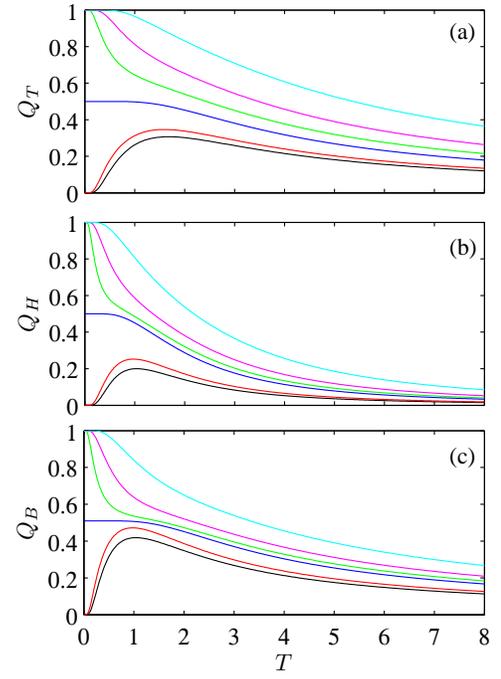}}
\caption{(Color online) $Q_T(\rho)$, $Q_H(\rho)$, and $Q_B(\rho)$
versus $T$ with $J=1$ and $B=1.5$. Here, the black, red, blue,
green, magenta, and cyan curves (from bottom to top) correspond to
the cases of $D=0$, 0.5, $\sqrt{5}/2$, 1.5, 2, and 3,
respectively.}\label{fig:4}
\end{figure}

Finally, we discuss the temperature dependence of the three
GQDs. We will show that while the increase of $T$
can in general destroy the coherence of the
system, the GQDs may also be increased by increasing
$T$ for certain specific system parameters. For this purpose, we showed in Fig.
\ref{fig:4} the $T$ dependence of $Q_T(\rho)$, $Q_H(\rho)$, and
$Q_B(\rho)$ with fixed $B=1.5$ and different DM interactions,
from which some general behaviors can be observed.

First, when $\delta<B$ (e.g., $D=0$ and $0.5$ shown in Fig.
\ref{fig:4}), $Q_T(\rho)$, $Q_H(\rho)$, and $Q_B(\rho)$ initially
increase from zero to certain maximum values, and then decrease to
zero gradually. Therefore, for this case the increase of $T$ can enhance
the GQDs to some extent. Second, when $\delta= B$
(e.g., $D=\sqrt{5}/2$ shown in Fig. \ref{fig:4}), the GQDs decrease
monotonously from the initial value 0.5 [for
$Q_T(\rho)$ and $Q_H(\rho)$] or 0.5098 [for $Q_B(\rho)$] in the
whole region of $T$, and arrives at the asymptotic value zero in the
infinite temperature limit. Thirdly, when $\delta>B$ (e.g., $D=1.5$,
2, and 3 shown in Fig. \ref{fig:4}), all the three GQDs decrease
from the maximum value 1 (when $T=0$) to the
minimum value 0 in the infinite temperature limit. Thus, one can see that
for the latter two cases, the increase of temperature always degrades
the GQDs.

\section{Summary}\label{sec:5}
In summary, we have investigated properties of the GQDs for two spins
in thermal equilibrium. The corresponding physical system we considered
is described by the Heisenberg {\it XX} model, with the DM interaction
induced by the spin-orbit coupling being involved, and an external
magnetic field is also applied. Moreover, the three GQDs we
considered were defined based on the trace distance, the Hellinger
distance, and the Bures distance, respectively. By analyzing their
dependence on the system parameters, we found that the DM interaction
plays a positive role in improving the GQDs. To be explicitly, we found
that the GQDs can be enhanced apparently by introducing the DM
interaction. Particularly, they all approach the maximum 1 in the
infinite limit of $D$.

On the other hand, the applied external magnetic fields always degrade
the trace distance GQD, while its effects on the Hellinger
distance and the Bures distance GQDs are temperature dependent.
For the low temperature case, $Q_H(\rho)$ is always decreased, while
$Q_B(\rho)$ may be enhanced in a narrow region of $B$. For the high
temperature case with weak DM interaction, both $Q_H(\rho)$ and
$Q_B(\rho)$ can also be enhanced in the weak magnetic field region.

Finally, the temperature dependence of the three GQDs are
determined by the relative magnitudes of $\delta$ and $B$. When
$\delta<B$, the GQDs may be enhanced to some extent by increasing
$T$ at the low temperature region, and when $\delta\geq B$, they
are degraded in the whole temperature region.

\newcommand{\PRL}{Phys. Rev. Lett. }
\newcommand{\RMP}{Rev. Mod. Phys. }
\newcommand{\PRA}{Phys. Rev. A }
\newcommand{\PRB}{Phys. Rev. B }
\newcommand{\PRE}{Phys. Rev. E }
\newcommand{\NJP}{New J. Phys. }
\newcommand{\JPA}{J. Phys. A }
\newcommand{\JPB}{J. Phys. B }
\newcommand{\OC}{Opt. Commun. }
\newcommand{\PLA}{Phys. Lett. A }
\newcommand{\EPJD}{Eur. Phys. J. D }
\newcommand{\NP}{Nat. Phys. }
\newcommand{\NC}{Nat. Commun. }
%


\begin{thebibliography}{50}

\bibitem{rmp-e}R. Horodecki, P. Horodecki, M. Horodecki, K. Horodecki, \RMP 81 (2009) 865.
\bibitem{qt1}C.H. Bennett, et al., \PRL 70 (1993) 1895.
\bibitem{qt2}M.L. Hu, H. Fan, \PRA 86 (2012) 032338.
\bibitem{disen}M.L. Hu, Ann. Phys. 327 (2012) 2332.
\bibitem{qd-en}H. Ollivier, W.H. Zurek, \PRL 88 (2001) 017901;\\
               L. Henderson, V. Vedral, \JPA 34 (2001) 6899.
\bibitem{rmp-qd}K. Modi, A. Brodutch, H. Cable, T. Paterek, V. Vedral, \RMP 84 (2012) 1655.
\bibitem{dqc1}A. Datta, A. Shaji, C.M. Caves, \PRL 100 (2008) 050502.
\bibitem{rsp}B. Daki\'{c}, et al., \NP 8 (2012) 666.
\bibitem{qlock1}A. Datta, S. Gharibian, \PRA 79 (2009) 042325.
\bibitem{qlock2}S. Wu, U.V. Poulsen, K. M{\o}lmer, \PRA 80 (2009) 032319.

\bibitem{Gumile}M. Gu {\it et al.}, \NP 8 (2012) 671.
\bibitem{cpm}A. Shabani, D.A. Lidar, \PRL 102 (2009) 100402.
\bibitem{eur1}M.L. Hu, H. Fan, \PRA 87 (2013) 022314.
\bibitem{eur2}M.L. Hu, H. Fan, \PRA 88 (2013) 014105.
\bibitem{mid}S. Luo, \PRA 77 (2008) 022301.

\bibitem{min1}S. Luo, S. Fu, \PRL 106 (2011) 120401;\\
              S. Luo, S. Fu, Europhys. Lett. 92 (2010) 20004.
\bibitem{min2}M.L. Hu, H. Fan, Ann. Phys. 327 (2012) 2343.
\bibitem{min3}Z. Xi, X. Wang, Y. Li, \PRA 85 (2012) 042325.
\bibitem{gqd-sch}B. Daki\'{c}, V. Vedral, \v{C}. Brukner, \PRL 105 (2010) 190502;\\
                 S. Luo, S. Fu, \PRA 82 (2010) 034302.
\bibitem{Problem}M. Piani, \PRA 86 (2012) 034101;\\
                 X. Hu, H. Fan, D.L. Zhou, W.M. Liu, \PRA 87 (2013) 032340.

\bibitem{trace}F.M. Paula, T.R. de Oliveira, M.S. Sarandy, \PRA 87 (2013) 064101;\\
               B. Aaronson, R.L. Franco, G. Compagno, G. Adesso, \NJP 15 (2013) 093022.
\bibitem{square}L. Chang, S. Luo, \PRA 87 (2013) 062303.
\bibitem{lqu}D. Girolami, T. Tufarelli, G. Adesso, \PRL 110 (2013) 240402.
\bibitem{bures}B. Aaronson, R.L. Franco, G. Adesso, \PRA 88 (2013) 012120;\\
               D. Spehner, M. Orszag, \NJP 15 (2013) 103001.
\bibitem{pres1}L. Mazzola, J. Piilo, S. Maniscalco, \PRL 104 (2010) 200401.

\bibitem{pres2}A. Streltsov, H. Kampermann, D. Bru{\ss}, \PRL 107 (2011) 170502.
\bibitem{pres3}M.L. Hu, D.P. Tian, Ann. Phys. 343 (2014) 132.
\bibitem{dyn1}T. Werlang, S. Souza, F.F. Fanchini, C.J. Villas Boas, \PRA 80 (2009) 024103.
\bibitem{dyn2}J. Maziero, T. Werlang, F.F. Fanchini, L.C. C\'{e}leri, R.M. Serra, \PRA 81 (2010) 022116.
\bibitem{dyn3}B. Wang, Z.Y. Xu, Z.Q. Chen, M. Feng, \PRA 81 (2010) 014101.

\bibitem{dyn4}F.F. Fanchini, T. Werlang, C.A. Brasil, L.G.E. Arruda, A.O. Caldeira, \PRA 81 (2010) 052107.
\bibitem{dyn5}K. Berrada, F.F. Fanchini, S. Abdel-Khalek, \PRA 85 (2012) 052315.
\bibitem{dyn6}B. Bellomo, R.L. Franco, G. Compagno, \PRA 86 (2012) 012312.
\bibitem{dyn7}M.L. Hu, H. Fan, Ann. Phys. 327 (2012) 851.
\bibitem{dyn-trace}J.D. Montealegre, F.M. Paula, A. Saguia, M.S. Sarandy, \PRA 87 (2013) 042115.

\bibitem{sp1}J. Maziero, H.C. Guzman, L.C. C\'{e}leri, M.S. Sarandy, R.M. Serra, \PRA 82 (2010) 012106.
\bibitem{sp2}L. Ciliberti, R. Rossignoli, N. Canosa, \PRA 82 (2010) 042316.
\bibitem{sp3}T. Werlang, G. Rigolin, \PRA 81 (2010) 044101.
\bibitem{sp4}L. Ciliberti, R. Rossignoli, N. Canosa, \PRA 82 (2010) 042316.
\bibitem{sp5}B.Q. Liu, B. Shao, J.G. Li, J. Zou, L.A. Wu, \PRA 83 (2011) 052112.

\bibitem{sp6}Y.C. Li, H.Q. Lin, \PRA 83 (2011) 052323.
\bibitem{sp7}G.F. Zhang, H. Fan, A.L. Ji, Z.T. Jiang, A. Abliz, W.M. Liu, Ann. Phys. 326 (2011) 2694.
\bibitem{sp8}C. Wang, Y.Y. Zhang, Q.H. Chen, \PRA 85 (2012) 052112.
\bibitem{qpt}T. Werlang, C. Trippe, G.A.P. Ribeiro, G. Rigolin, \PRL 105 (2010) 095702.
\bibitem{qpt1}R. Dillenschneider, \PRB 78 (2008) 224413.

\bibitem{qpt2}M.S. Sarandy, \PRA 80 (2009) 022108.
\bibitem{qpt3}F. Altintas, R. Eryigit, Ann. Phys. 327 (2012) 3084.
\bibitem{DM1}G.F. Zhang, \PRA 75 (2007) 034304.
\bibitem{DM2}M.L. Hu, \PLA 374 (2010) 3520.
\bibitem{ana-trace}F. Ciccarello, T. Tufarelli, V. Giovannetti, \NJP 16 (2014) 013038.

\bibitem{ana-bures}D. Spehner, M. Orszag, \JPA 47 (2014) 035302.



\end{thebibliography}
%

\end{document}